\documentclass[aps,prl,oneside,twocolumn]{revtex4-1}
\usepackage{amsmath}
\usepackage{amssymb}
\usepackage{graphicx}
\usepackage{multirow}
\usepackage{booktabs}
\usepackage{mathtools}

\begin{document}
\draft \title{The second hyperpolarizability of systems described by the space-fractional Schr\"{o}dinger equation}
\author{Nathan J. Dawson}
\email{dawsphys@hotmail.com}
\address{Department of Natural Sciences, Hawaii Pacific University, Kaneohe, HI, 96744, USA}
\address{Department of Computer Science and Engineering, Hawaii Pacific University, Honolulu, HI 96813, USA}
\author{Onassis Nottage}
\address{Department of Physics, University of The Bahamas, Nassau, Bahamas}
\author{Moussa Kounta}
\address{Department of Mathematics, University of The Bahamas, Nassau, Bahamas}

\begin{abstract}
The static second hyperpolarizability is derived from the space-fractional Schr\"{o}dinger equation in the particle-centric view. The Thomas-Reiche-Kuhn sum rule matrix elements and the three-level ansatz determines the maximum second hyperpolarizability for a space-fractional quantum system. The total oscillator strength is shown to decrease as the space-fractional parameter $\alpha$ decreases, which reduces the optical response of a quantum system in the presence of an external field. This damped response is caused by the wavefunction dependent position and momentum commutation relation. Although the maximum response is damped, we show that the one-dimensional quantum harmonic oscillator is no longer a linear system for $\alpha \neq 1$, where the second hyperpolarizability becomes negative before ultimately damping to zero at the lower fractional limit of $\alpha \rightarrow 1/2$.
\end{abstract}

\maketitle

\section{Introduction}
\label{intro}

Kuzyk first discovered limits to the nonlinear optical responses of non-relativistic systems with position dependent potentials.\cite{kuzyk00.01} These limits are much greater than the largest responses obtained through experimentation.\cite{kuzyk13.01} Another gap has also been observed between the fundamental limits and the best optimized pseudo-potentials.\cite{ather12.01,burke13.01,dawson16.02,watki12.01} These reported gaps may be better understood by investigating more generalized quantum mechanical theories. The relativistically corrected Thomas-Reiche-Kuhn (TRK) sum rule \cite{cohen98.01,sinky06.01} led to smaller intrinsic nonlinearities as compared to those calculated in the purely non-relativistic regime, where this decrease in the response is caused by higher-order momentum operators appearing from block diagonalization of the Dirac equation.\cite{dawson15.01} The first hyperpolarizability of systems described by the space-fractional Schr\"{o}dinger equation has also been investigated.\cite{dawson16.01}

Laskin discovered the space-fractional Schr\"{o}dinger equation by generalizing the path integral formulation using a L\'{e}vy-type path.\cite{laski00.01} Laskin further investigated the space-fractional Schr\"{o}dinger equation, where he formulated a fractional generalization of the Heisenberg uncertainty principal, proved the Hermiticity of the fractional Hamiltonian operator, and determined the energy spectrum of space-fractional, hydrogen-like atoms.\cite{laski00.02,laski02.01} The kinetic energy in the space-fractional Schr\"{o}dinger equation depends on fractional momentum operators, which results in a fractional derivative. The Riesz fractional derivative \cite{riesz49.01} $\left(-\nabla^2\right)^\alpha$ appears in the space-fractional Schr\"{o}dinger equation.

In this paper, we derive a sum-over-states expression for the second hyperpolarizability. The limit to the second hyperpolarizability from the space-fractional Schr\"{o}dinger equation depends on the fractional parameter $\alpha$, therefore we define an apparent intrinsic second hyperpolarizability to make comparisons between the space-fractional Schr\"{o}dinger equation and the standard Schr\"{o}dinger equation. Although the limit to the second hyperpolarizability decreases when $\alpha$ is reduced below unity, we show that some potentials with a small nonlinear optical response can gain a larger response magnitude. This is explicitly shown for the quantum harmonic oscillator, which has a non-zero second hyperpolarizability determined from the fractional Schr\"{o}dinger equation within the L\'{e}vy index $1 < 2\alpha \leq 2$.

\section{Theory}
\label{sec:1}

The time-independent space-fractional Schr\"{o}dinger equation with a momentum operator given by the Riesz fractional derivative for a single particle system is given by
\begin{equation}
\hat{H}_\alpha \psi = E \psi ,
\label{eq:fracschrod}
\end{equation}
where $\hat{H}_\alpha$ is the space-fractional Hamiltonian with fractional parameter $1/2 < \alpha \leq 1$, $E$ is the energy, and $\psi$ is the wavefunction. The Hamiltonian considered in this paper has a kinetic energy described by the fractional momentum operator and a spatially dependent potential. The one-dimensional, space-fractional Hamiltonian may be written as
\begin{equation}
\hat{H}_{\alpha} = \frac{\hat{p}^2}{2m} + V\left(\hat{x}\right).
\label{eq:hamilt}
\end{equation}
where $m$ is the rest mass and $V\left(\hat{x}\right)$ is the potential energy.

Respectively, the position and momentum operators are given by
\begin{equation}
\hat{x} = \left(\frac{\hbar}{mc}\right)^{1-\alpha} \left|x\right|^\alpha \mathrm{sign}\left(x\right)
\label{eq:canonpos}
\end{equation}
and
\begin{equation}
\hat{p} = -i mc \left(\frac{\hbar}{mc}\right)^{\alpha} \frac{\partial^\alpha}{\partial x^\alpha} .
\label{eq:canonmom}
\end{equation}
The operator $\partial^\alpha/\partial x^\alpha$ in Eq. \ref{eq:canonmom} is a fractional derivative. There are many definitions of the fractional derivative; simulations performed in this paper are based on a numerical approximation to the Riesz fractional derivative. Note that the dimensions of linear space and momentum are preserved by the constants in Eqs. \ref{eq:canonpos} and \ref{eq:canonmom}, where $c$ is the speed of light in vacuum and $\hbar$ is the reduced Planck constant.

We use time-independent perturbation theory of the space-fractional Schr\"{o}dinger equation in one dimension to determine the scalar, static, second hyperpolarizability in the zero frequency limit.\cite{boyd08.01} The perturbing term in the Hamiltonian caused by the constant electric field ${\cal E}$ is given by
\begin{equation}
\hat{H}_{\alpha}^{\mathrm{pert}} = e {\cal E} \hat{x} ,
\label{eq:perthamilt}
\end{equation}
where $\hat{H}_\alpha = \hat{H}_{\alpha}^{\left(0\right)} + \hat{H}_{\alpha}^{\mathrm{pert}}$ with $\hat{H}_{\alpha}^{\left(0\right)}$ given by Eq. \ref{eq:hamilt}. We take a particle-centric approach, where the origin is placed at the expectation value of an electron in a potential well. Note that we may remove subscripts for single electron systems, where multi-electron systems will have different position operators based on the relative displacements between their origins at their respective expectation values. Only for $\alpha \rightarrow 1$ does the position operator and perturbation potential become linear.

The fourth-order correction to the energy from time-independent perturbation theory \cite{sakur94.01} is given by
\begin{align}
E^{\left(4\right)} &= \displaystyle\left. \sum_{k,\ell,n}\right.^\prime  \frac{\left(\hat{H}_{\alpha}^{\mathrm{pert}} \right)_{0k} \left(\overline{H}_{\alpha}^{\mathrm{pert}} \right)_{k \ell} \left(\overline{H}_{\alpha}^{\mathrm{pert}} \right)_{\ell n} \left(\hat{H}_{\alpha}^{\mathrm{pert}} \right)_{n 0}}{E_{k0} E_{\ell0} E_{n0}} \nonumber \\
&- \displaystyle\left. \sum_{k,\ell}\right.^\prime  \frac{\left(\hat{H}_{\alpha}^{\mathrm{pert}} \right)_{0k} \left(\hat{H}_{\alpha}^{\mathrm{pert}} \right)_{k 0} \left(\hat{H}_{\alpha}^{\mathrm{pert}} \right)_{ 0\ell} \left(\hat{H}_{\alpha}^{\mathrm{pert}} \right)_{\ell 0}}{E_{k0}^2 E_{\ell0}} ,
\label{eq:4thcorr}
\end{align}
where the prime denotes the sum over all states \textit{except} the ground state. Shorthand notation was introduced in Eq. \ref{eq:4thcorr}, $E_{ij} = E_{i}^{\left(0\right)} - E_{j}^{\left(0\right)}$ and $\overline{{\cal O}}_{ij} = \hat{{\cal O}}_{ij} - \delta_{ij} \hat{{\cal O}}_{00}$ with $\delta$ representing the Kronecker delta function, where $\hat{{\cal O}}_{ij} = \left\langle i^{\left(0\right)} \right| \hat{{\cal O}} \left| j^{\left(0\right)} \right\rangle$ is the transition probability of the unperturbed system with $\left|i^{\left(0\right)}\right\rangle$ being the unperturbed state vector indexed from the ground state $i=0$.

The static, third-order, scalar response is given by
\begin{equation}
\kappa^{\left(3\right)}= \displaystyle \frac{1}{\left(3\right)!} \displaystyle \left. \frac{\partial^4}{\partial {\cal E}^4} E_0\left({\cal E}\right) \right|_{{\cal E}=0} ,
\label{eq:kappan}
\end{equation}
where $E_0$ is the ground state energy. Thus, the sum-over-states expression for the static, scalar, second hyperpolarizability given in terms of the transition energies and fractional transition moments is
\begin{align}
\kappa^{\left(3\right)} = 4 \displaystyle e^4 \displaystyle \left.\sum_{k,\ell,n} \right.^\prime \frac{\hat{x}_{0k} \overline{x}_{k \ell} \overline{x}_{\ell n}\hat{x}_{n 0}}{E_{k0} E_{\ell 0} E_{n0}} - 4 \displaystyle e^4 \displaystyle \left.\sum_{k,\ell} \right.^\prime \frac{\hat{x}_{0k} \hat{x}_{k 0} \hat{x}_{0 \ell} \hat{x}_{\ell 0}}{E_{k0}^2 E_{\ell 0}} .
\label{eq:2ndhyperpol}
\end{align}
Because the theory is strictly particle-centric, the expectation value for an electron in its lowest energy state is always zero which allows us to neglect the bar operator in Eq. \ref{eq:2ndhyperpol}.

The Leibniz rule and chain rule known from integer calculus do not take the same form in fractional calculus, and therefore $\left[\hat{x},\hat{p}\right]$ will not, in general, be equal to the constant $i\hbar$ when $\alpha \neq 1$. The TRK sum rule \cite{thoma25.01,reich25.01,kuhn25.01} for the mechanical Hamiltonian found in the fractional Schr\"{o}dinger equation results in a wavefunction-dependent form. For a single electron, the fractional TRK sum rule, calculated from the transition probability of the second commutation relation of the Hamiltonian with the position operator $\left\langle k^{\left(0\right)} \right| [\hat{x},[\hat{H}_{\alpha}^{\left(0\right)}, x]] \left| \ell^{\left(0\right)} \right\rangle$, follows as
\begin{align}
\displaystyle \sum_{q = 0}^\infty \hat{x}_{kq} \hat{x}_{q\ell} \left[E_{q}^{\left(0\right)} - \frac{1}{2} \left(E_{k}^{\left(0\right)} + E_{\ell}^{\left(0\right)} \right)\right] = \displaystyle \frac{\hbar^2}{2 m} \, \lambda_\alpha\left(k,\ell\right) ,
\label{eq:TRK}
\end{align}
where
\begin{align}
\lambda_\alpha\left(k,\ell\right) &= \displaystyle \int \psi_{k}^{\left(0\right)\dag}\left(x\right) \bigg[\frac{1}{2}\hat{\xi}^2 \left(x\right) \frac{\partial^{2\alpha}}{\partial x^{2\alpha}} + \frac{1}{2}\frac{\partial^{2\alpha}}{\partial x^{2\alpha}}  \hat{\xi}^2\left(x\right) \nonumber \\
&- \hat{\xi}\left(x\right) \frac{\partial^{2\alpha}}{\partial x^{2\alpha}} \hat{\xi}\left(x\right) \bigg] \psi_{\ell}^{\left(0\right)}\left(x\right) \,dx
\label{eq:lambdaparam}
\end{align}
with $\psi_{i}^{\left(0\right)} \left(x\right) = \left\langle x | i^{\left(0\right)} \right\rangle$ and $\hat{\xi}\left(x\right) = \left|x\right|^\alpha \mathrm{sign}\left(x\right)$. The normalized wavefunction of the unperturbed system has the usual property,
\begin{equation}
\delta_{k\ell} = \displaystyle \int_{-\infty}^{\infty} \psi_{k}^{\left(0\right)\dag}\left(x\right) \psi_{\ell}^{\left(0\right)}\left(x\right)\, dx\, .
\label{eq:normcond}
\end{equation}
Note that the summation over the state $q$ is introduced into Eq. \ref{eq:lambdaparam} through the use of closure.

The $\left(k=0,\ell=0\right)$ TRK sum rule element gives,
\begin{equation}
E_{10}\left|\left(\hat{x}\right)_{10}\right|^2 = \frac{\hbar^2}{2 m} \lambda_\alpha\left(0,0\right)  - \displaystyle \sum_{q = 2}^{\infty} E_{q0}\left|\left(\hat{x}\right)_{q0}\right|^2 .
\label{eq:00TRKPre}
\end{equation}
It is clear from Eq. \ref{eq:00TRKPre} that the largest possible ground state transition moment allowed by the TRK sum rule happens when all of the oscillator strength is in the transition to the first excited state. Setting all terms in the sum for $q \geq 2$ equal to zero gives the maximum value of the ground state to first excited state transition moment,
\begin{equation}
\hat{x}_{10}^{\mathrm{max}} = \frac{\hbar }{\sqrt{2 m E_{10}}} \sqrt{\lambda_\alpha\left(0,0\right)} ,
\label{eq:xmax}
\end{equation}
where transition moments of a bound electron described by the space-fractional Schr\"{o}dinger equation with the Riesz fractional derivative and mechanical Hamiltonian are real, and therefore, $\hat{x}_{ij} = \hat{x}_{ji}$.

\begin{table}[t]
\caption{Fractional transition dipole moments for a three-level model as a function of $\hat{X}$, $E$, and $\lambda_\alpha\left(k,\ell\right)$}
\begin{tabular}{ r l }
\hline
Source & Transition dipole moment \\[5pt]
\hline
Eq. \ref{eq:Xdefine} & $\displaystyle\hat{x}_{10} = \frac{\hbar}{\sqrt{2 m E_{10}}} \hat{X} \sqrt{\lambda_\alpha\left(0,0\right)}$  \\[12pt]
TRK $(0,0)$ & $\displaystyle \hat{x}_{20} = \frac{\hbar}{\sqrt{2 m E_{10}}} \sqrt{E\left(1-\hat{X}^2\right)} \sqrt{\lambda_\alpha\left(0,0\right)}$ \\[12pt]
TRK $(1,1)$ & $\displaystyle \hat{x}_{12} = \frac{\hbar}{\sqrt{2 m E_{10}}} \sqrt{\frac{E}{1-E}}\sqrt{\hat{X}^2 \lambda_\alpha\left(0,0\right) + \lambda_\alpha\left(1,1\right)}$ \\[12pt]
TRK $(0,1)$ & $\displaystyle \overline{x}_{11} = \frac{\hbar}{\sqrt{2 m E_{10}}}\Bigg[ \frac{E-2}{\sqrt{1-E}} \frac{\sqrt{1-\hat{X}^2}}{\hat{X}}$ \\[12pt]
 & $\times\sqrt{\hat{X}^2 \lambda_\alpha\left(0,0\right) + \lambda_\alpha\left(1,1\right)}$ \\[12pt]
 & \hspace{0.5cm} $\displaystyle - \frac{1}{\hat{X}}\frac{\lambda_\alpha\left(1,0\right)}{\sqrt{\lambda_\alpha\left(0,0\right)}}\Bigg]$ \\[12pt]
TRK $(0,2)$ & $\displaystyle \overline{x}_{22} = \frac{\hbar}{\sqrt{2 m E_{10}}}\Bigg[ \frac{1-2E}{\sqrt{1-E}} \frac{\hat{X}}{\sqrt{1-\hat{X}^2}}$ \\[12pt]
 & $\times\sqrt{\hat{X}^2 \lambda_\alpha\left(0,0\right) + \lambda_\alpha\left(1,1\right)}$ \\[12pt]
 & \hspace{0.5cm}$\displaystyle - \sqrt{\frac{E}{1-\hat{X}^2}} \frac{\lambda_\alpha\left(2,0\right)}{\sqrt{\lambda_\alpha\left(0,0\right)}}\Bigg]$ \\[12pt]
\hline
\end{tabular}
\label{table:transmom}
\end{table}

The maximum hyperpolarizability derived from the TRK sum rule with only three levels has traditionally been regarded as the fundamental limit. The three-level ansatz appears to hold when the response is near the fundamental limit for a mechanical Hamiltonian in the standard Schr\"{o}dinger equation. For the case of the fractional Schr\"{o}dinger equation, the fractional TRK sum rule gives a reduced value which lowers the limit while the transition dipole moment and energy eigenvalue dependencies are of the same form as the TRK sum rule derived for the standard case. Thus, we expect the three-level ansatz to hold for the systems described by the space-fractional Schr\"{o}dinger equation.

The fundamental limit to the scalar second hyperpolarizability derived from the standard Schr\"{o}dinger equation with the three-level ansatz is well-established, $\gamma_\mathrm{max} = 4 e^4 \hbar^4 / m^2 E_{10}^{5}$.\cite{perez08.01} Using the three-level ansatz, we define the parameters
\begin{equation}
E = E_{10}/E_{20}
\label{eq:Edefine}
\end{equation}
and
\begin{equation}
\hat{X} = \left|\hat{x}_{10}\right| / \hat{x}_{10}^{\mathrm{max}} \, .
\label{eq:Xdefine}
\end{equation}
Multiplying both sides of Eq. \ref{eq:Xdefine} by $\hat{x}_{10}^{\mathrm{max}}$ gives the first transition moment relationship in Table \ref{table:transmom}.

All transition dipole moments for a three-level model of the space-fractional Schr\"{o}dinger equation with a mechanical Hamiltonian can be expressed in terms of $\hat{X}$, $E$, and $\lambda_\alpha\left(k,\ell\right)$. These remaining transition dipole moments are given in Table \ref{table:transmom}. Substituting the transition moment expressions from Table \ref{table:transmom} into Eq. \ref{eq:2ndhyperpol}, the three-level, space-fractional, second hyperpolarizability becomes
\begin{align}
\kappa^{\left(3\right)} &= \frac{e^4 \hbar ^4}{m^2 E_{10}^5} \Bigg\{\bigg[2 \hat{X}^2 \left(1-E^2\right) \left(2-E^3\right)-5 \hat{X}^4 (1-E) \nonumber \\
&\times \left(1-E^2\right) \left(1+E+E^2\right)-E^5\bigg] \lambda_{\alpha}^2\left(0,0\right) \nonumber \\ &+(1-E) \left[(2+E)^2-4 \hat{X}^2 \left(1+E-E^3-E^4\right)\right] \nonumber \\
&\times \lambda_{\alpha}\left(0,0\right) \lambda_{\alpha}\left(1,1\right)+\lambda_{\alpha}^2\left(1,0\right) \nonumber \\
&+\sqrt{(1-E)} \sqrt{\lambda_{\alpha}\left(0,0\right) \left(\hat{X}^2 \lambda_{\alpha}\left(0,0\right)+\lambda_{\alpha}\left(1,1\right)\right)} \nonumber \\
&\times \bigg[2 \hat{X} E^{7/2} \left(1+2 E\right) \lambda_{\alpha}\left(2,0\right) \nonumber \\
&-2 \sqrt{\left(1-X^2\right)}\left(2+E\right) \lambda_{\alpha}\left(1,0\right) \bigg]+E^5 \lambda_{\alpha}^2\left(2,0\right)\Bigg\} .
\label{eq:3Lhyper}
\end{align}

When approaching integer dimensions, \textit{i}.\textit{e}., $\alpha\rightarrow 1$, the fractional $\lambda_\alpha$ coefficients reduce to $\lambda_{\alpha\rightarrow1}\left(i,i\right)\rightarrow 1$ and $\lambda_{\alpha\rightarrow1}\left(i,j\right)\rightarrow 0$ for $i\neq j$. Thus, the space-fractional Schr\"{o}dinger equation is reduced to the standard Schr\"{o}dinger equation for the case of $\alpha\rightarrow 1$ along with the Liebniz rule and chain rule from integer calculus, which returns the standard TRK sum rule. The maximum second hyperpolarizability of a bound particle of charge $e$ for the case of $\alpha\rightarrow 1$ occurs when $\hat{X} = 0$ and $E = 0$,
\begin{equation}
\kappa_{\mathrm{max},\alpha\rightarrow 1}^{\left(3\right)} = 4\frac{e^4 \hbar^4}{m^2 E_{10}^{5}} .
\label{eq:3Lhypermax}
\end{equation}

As $\alpha$ is reduced below unity, the diagonal elements $\lambda_\alpha\left(0,0\right)$ and $\lambda_\alpha\left(1,1\right)$ can fall below unity. Thus, the reduced order of the momentum operator in the kinetic energy term reduces the fundamental limit of the second hyperpolarizability. The terms that contain the parameters $\lambda_\alpha\left(1,0\right)$ and $\lambda_\alpha\left(2,0\right)$ in Eq. \ref{eq:3Lhyper} are consistently near zero for every class of potential that we have numerically evaluated when in the particle centric view; however, these terms become large non-zero values when the origin is moved away from the ground state expectation value. The TRK sum rule is not wavefunction dependent for $\alpha = 1$, which is a known result for the TRK sum rule derived from standard Schr\"{o}dinger equation with a mechanical Hamiltonian.

\section{Discussion}

\begin{figure}[t!]
\centering\includegraphics[scale=0.8]{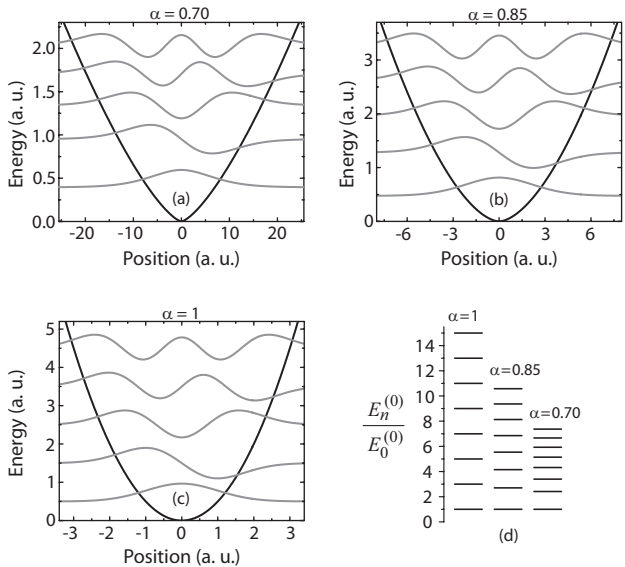}
\caption{The potential well (black line) along with the ground state and first four excited state wavefunctions (gray lines) for a fractional QHO. The single bound particle systems are shown for (a) $\alpha = 0.70$, (b) $\alpha = 0.85$, and (c) $\alpha = 1$. (d) The first 8 energy eigenvalues for the fractional QHO for the $\alpha$ values used to generate graphs (a), (b), and (c).}
\label{fig:FracQHO}
\end{figure}

The fundamental limit of the second hyperpolarizability occurs at the integer dimension limit. For fixed $\alpha \neq 1$, the diagonal elements of the TRK sum rule depend on the wavefunction. Thus, the fundamental limit at $\alpha \neq 1$ depends on the potential. Therefore, we define the apparent intrinsic second hyperpolarizability,
\begin{equation}
\kappa_{\mathrm{app}}^{\left(3\right)} = \frac{\kappa^{\left(3\right)}}{\kappa_{\mathrm{max},\alpha\rightarrow1}^{\left(3\right)}} ,
\label{eq:appint}
\end{equation}
which compares the space-fractional second hyperpolarizability to the maximum second hyperpolarizability when $\alpha = 1$.

The fractional quantum harmonic oscillator (QHO) is defined with the potential
\begin{equation}
V\left(\hat{x}\right) = \frac{1}{2} m \omega^2 \hat{x}^2
\label{eq:CQHO}
\end{equation}
where $\omega$ is the angular frequency and the origin is placed at the ground state expectation value, $\hat{x}_{00} = 0$. The fractional QHO in integer space is a linear system. Thus, the second hyperpolarizability is zero for $\alpha = 1$.

The potential and first four wavefunctions of the fractional QHO are shown in Fig. \ref{fig:FracQHO} for (a) $\alpha = 0.70$, (b) $\alpha = 0.85$, and $\alpha = 1$, which are given in atomic units (a.$\,$u.). It is obvious that centrosymmetry is preserved when choosing the origin at the minimum of the potential well. Figure \ref{fig:FracQHO}(d) shows the change in the spacing between transitions as $\alpha$ is decreased. In addition to the unevenly spaced energy spectrum of the fractional QHO for $0 < \alpha < 1$, the entirety of the oscillator strength is no longer in the ground to first excited state transition and allows for a nonzero second hyperpolarizability.

We used a finite-difference approximation of the Riesz fractional derivative \cite{podlu09.01} via the half sum of the left- and right-sided Caputo fractional derivatives to numerically determine the energy eigenvalues and wavefunctions. Note that the central difference scheme converges to the usual local second-order finite-difference approximation when $\alpha \rightarrow 1$; however, the scheme is nonlocal for $\alpha \neq 1$, which made sparse matrix calculations more time consuming for decreasing $\alpha$. Thus, we used Cholesky factorization to determine the energy levels and wavefunctions. The calculations shown in Fig. \ref{fig:FracQHO} are approximated with an $8000\times8000$ Hamiltonian matrix, where each Dirichlet boundary is set far from the potential so that the wave functions converge near zero for a significant fraction of the total calculated domain.

\begin{figure}[t]
\centering\includegraphics[scale=0.7]{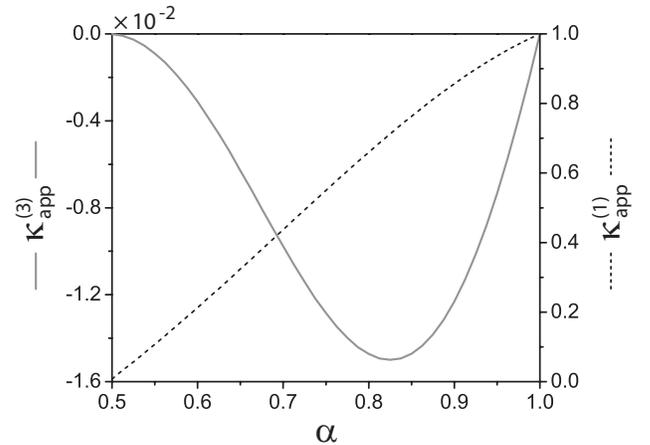}
\caption{The apparent intrinsic polarizability $\kappa_{\mathrm{app}}^{\left(1\right)}$ and apparent intrinsic second hyperpolarizability $\kappa_{\mathrm{app}}^{\left(2\right)}$ for the fractional QHO ($\omega = 1$ a.u.) as a function $\alpha$.}
\label{fig:fracvals}
\end{figure}

The apparent intrinsic polarizability and apparent intrinsic second hyperpolarizability of the fractional QHO as a function of $\alpha$ are shown in Fig. \ref{fig:fracvals}. The apparent intrinsic second hyperpolarizability is calculated using the first $20$ eigenstates with the sum-over-states expression given in Eq. \ref{eq:2ndhyperpol}. It was previously discovered that the hyperpolarizability was more sensitive to the value of $\alpha$ than the polarizability, where $\kappa^{\left(1\right)}$ is of order $\lambda_\alpha$ while $\kappa^{\left(2\right)}$ is of order $\lambda_{\alpha}^{3/2}$. The second hyperpolarizability is proportional to a sum of four multiplicative fractional dipole transition moments, which is of order $\lambda_{\alpha}^{2}$. Thus, we expect that the magnitude of the space-fractional second hyperpolarizability to decrease faster than the polarizability and hyperpolarizability with decreasing $\alpha$.

It was previously noted that moving the origin away from the particle-centric position, $\hat{x}_{00}$, caused the hyperpolarizability to further decrease when $\alpha \neq 1$. Regarding broad physical and mathematical concerns, centrosymmetric systems have an even more interesting consequence for moving away from the particle-centric model. When the origin of the space-fractional system is not located at $\hat{x}_{00}$, a centrosymmetric potential becomes asymmetric for $\alpha \neq 1$ and the hyperpolarizability is no longer zero. Thus, although the maximum possible second hyperpolarizability is reduced when $\alpha$ is taken just below unity, the hyperpolarizability can be increased for this general case. This type of symmetry breaking is not a concern when we constrain ourselves to the particle-centric view.

\begin{figure}[t]
\centering\includegraphics[scale=0.75]{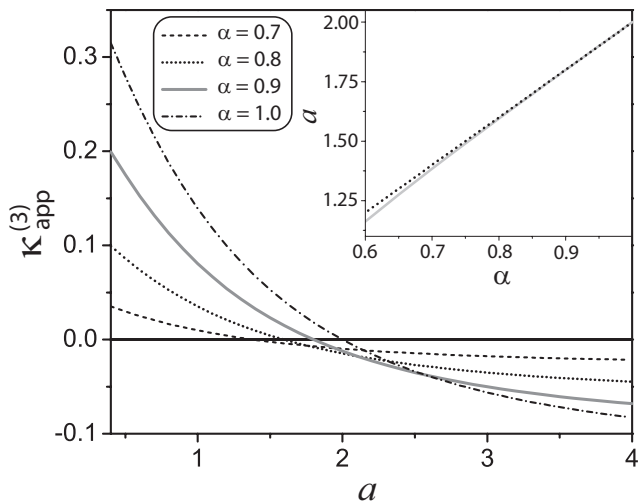}
\caption{The apparent intrinsic second hyperpolarizability as a function of the power, $a$, of a symmetric power potential with $b=1\,$a.u. The inset shows the power, $a$, as a function of $\alpha$ when the second hyperpolarizability is zero, where the dotted line is numerically approximated and the gray line is the theoretical slope of 2.}
\label{fig:power}
\end{figure}

An interesting consequence of relaxing our notion of integer operators is discovered while investigating the fractional QHO, where a linear system becomes nonlinear with a negative fractional second hyperpolarizability. As the fractional parameter is further reduced, the drop in the net oscillator strength dominates the response, where $\kappa^{\left(1\right)} \rightarrow 0$ as $\alpha \rightarrow 1/2$. Examining the three-level model given by Eq. \ref{eq:3Lhyper}, each term has a $\lambda_{\alpha}^2$ dependence. Evaluating Eq. \ref{eq:lambdaparam} for any wavefunction satisfying the fractional Schr\"{o}dinger equation when $\alpha \rightarrow 1/2$ gives a zero value for the diagonal $\lambda_\alpha$ matrix. The off-diagonal coefficients of the $\lambda_\alpha$ matrix are zero in the particle centric view. Thus, it is of no surprise that the second hyperpolarizability approaches zero as $\alpha \rightarrow 1/2$. Note that the three-level model of the first hyperpolarizability has a $\lambda_{\alpha}^{3/2}$ dependence and the polarizability has a linear dependence on $\lambda_\alpha$, which fall to zero as $\alpha \rightarrow 1/2$ at a lower rate than the second hyperpolarizability.

The consequence of transforming a linear system such as the QHO into a nonlinear system warrants further investigation. There is a slow deviation from linearity as $\alpha$ is decreased from unity, therefore we suspect that there exists a symmetric power potential that constitutes a linear system for $\alpha \neq 1$,
\begin{equation}
\displaystyle V = b \left|\hat{x}\right|^a ,
\label{eq:sympowpot}
\end{equation}
where $b = 1\,$a.u. is chosen for simplicity.

The main plot in Fig. \ref{fig:power} shows the apparent intrinsic second hyperpolarizability as a function of the power, $a$, for different values of $\alpha$. The second hyperpolarizability shows the same trend for all $\alpha$ but with differing magnitudes and axis crossings. As $\alpha$ is decreased, the power $a$ also decreases for a potential with $\kappa_{\mathrm{app}}^{\left(3\right)} = 0$ corresponding to a zero second hyperpolarizability. This zero second hyperpolarizability is caused by the entire oscillator strength being in the ground state to first excited state transition. Thus, the first summation on the right-hand-side of Eq. \ref{eq:2ndhyperpol} perfectly cancels with the second summation.

The inset in Fig. \ref{fig:power} shows $a$ as a function of $\alpha$ when the second hyperpolarizability of the symmetric power potential is zero. The dotted line represents the numerical approximation of the fractional Schr\"{o}dinger equation and minimization of the apparent intrinsic second hyperpolarizability and the gray line represents the theoretically predicted line with a slope of 2. The disagreement between the two lines representing the numerical approximation and the theoretical prediction are from the limitation of a $3000\times3000$ matrix representing the Hamiltonian in the numerical approximation as well as the increased shallowness of the minimum of $\left|\kappa_{\mathrm{app}}^{\left(3\right)}\right|$ as $\alpha$ becomes small. The increased shallowness in the region near the minimum of $\left|\kappa_{\mathrm{app}}^{\left(3\right)}\right|$ is observed in the main plot of Fig. \ref{fig:power} by the reduced angle when crossing the axis at small $\alpha$.

\section{Conclusion}

The static, space-fractional, second hyperpolarizability was derived from the space-fractional Schr\"{o}dinger equation. The space-fractional TRK sum rule elements were shown to be equal to a wavefunction dependent quantity after examining $\left[\hat{x},\left[\hat{H}_{\alpha}^{\left(0\right)},\hat{x}\right]\right]$ for $1/2<\alpha<1$. The three-level model contained two diagonal $\lambda_\alpha$ parameters. The two off-diagonal $\lambda_\alpha$ were zero for symmetric power law functions, which is in line with previous findings in that these same off-diagonal terms were nearly zero for numerically evaluated asymmetric potentials with the origin fixed at $\hat{x}_{00}$.\cite{dawson16.01} The generalized expression for the three-level model is reduced to that derived from the standard Schr\"{o}dinger equation with a mechanical Hamiltonian when setting $\lambda_\alpha\left(i,i\right) = 1$ and $\lambda_\alpha\left(i,j\right) = 0$ for $i\neq j$.

The apparent intrinsic second hyperpolarizability was presented as the space-fractional quantum system's second hyperpolarizability divided by the maximum second hyperpolarizability allowed by the TRK sum rule elements derived from the standard Schr\"{o}dinger equation. The fractional QHO was shown to change from a linear to a nonlinear system, where the second hyperpolarizability became negative. The linear and nonlinear response approached zero when $\alpha \rightarrow 1/2$ as expected. We also showed that for $1/2<\alpha\leq 1$, there exists a symmetric power potential that corresponds to a linear system with $\kappa_{\mathrm{app}}^{\left(3\right)} = 0$. The trendline for $a$ as a function of $\alpha$ as compared to the theoretical slope of 2 was shown to be a useful metric for the accuracy of the numerical approximations to the space-fractional Schr\"{o}dinger equation.

Although the non-relativistic quantum mechanical description of spinless, charged, quantum systems can be described by the standard Schr\"{o}dinger equation, there may be some quasi-particle, exotic-particles, and long-range systems with a space-fractional Schr\"{o}dinger equation description. There has also been some recent success identifying optical systems that are described by the space-fractional Schr\"{o}dinger equation.\cite{longhi15.01,zhang16.01} Thus, we might also anticipate some quantum systems with complicated charged particle dynamics that can be approximated by the space-fractional Schr\"{o}dinger equation. Wei \cite{wei15.01} pointed out that when $\alpha \rightarrow 1/2$, the kinetic energy of the space-fractional Schr\"{o}dinger equation approaches the same order of momentum dependence as the kinetic energy in the relativistic Schr\"{o}dinger equation. Because the net oscillator strength approaches zero for bound systems of charged particles as $\alpha \rightarrow 1/2$, we expect approximate low $\alpha$ space-fractional Schr\"{o}dinger equation systems and approximate relativistic Schr\"{o}dinger equation systems to interact weakly with photons.




\end{document}